\documentclass[a4paper,11pt,1,5 lines]{article}
\usepackage{amssymb,amsmath}
\usepackage{amsfonts}
\usepackage{indentfirst}
\usepackage{graphicx}
\newcommand{\ds}{\displaystyle}
\title{ Shifted polynomials in a convection problem}
\author{ Ioana Dragomirescu\\
\\
Department of Mathematics,
 University "Politehnica" of Timisoara,\\
 ioana.dragomirescu@mat.upt.ro}
\begin{document}
\date{}
\maketitle \begin{abstract} The onset of convection in a horizontal layer of fluid heated from below in the presence of a gravity field varying
across the layer is investigated. The eigenvalue problem governing the linear stability of the mechanical equilibria of the fluid layer in the case of
free boundaries is solved using a Galerkin method based on shifted polynomials (Legendre and Chebyshev polynomials).
\end{abstract}
\section{Statement of the problem}
Physical problems concerning the motion of fluids in the presence of a variable gravity field can be encountered in many practical applications, e.g.
convection problems in porous media, crystal growth domain or problems concerning the mass transport in the Earth's system. That is way,
convection problems with variable gravity fields have been intensively studied. In this paper we analyze the influence of such gravity
fields varying across the layer on the stability bounds in a convection problem that arises in a horizontal layer of fluid heated from below.
The gravity field acting in the $z$-direction is orthogonal to the fluid layer and is assumed to depend on the vertical
coordinate $z$ only \cite{Str}. For such a variable gravity field different points of the fluid experience different buoyancy forces. As a
consequence, part of a fluid layer tends to become unstable while the other tends to remain stable, the mechanical equilibrium turning into a
convective motion.
\par Experimental measurements of the Earth's upper atmosphere
show that the atmospheric density decreases as the altitude increases as an approximately exponentially function of the vertical height.
\par Taking these into account, consider a layer of heat-conducting viscous fluid
contained between the planes $z=0$ and $z=h$\cite{Str}. The equations governing  the convective motion and the conducting state are [8]
\begin{equation}
\left\{
\begin{array}{l}
\dfrac{\partial {\bf v}}{\partial t}+({\bf v}\cdot \textrm{grad}){\bf v}=-\frac{1}{\rho}\textrm{grad}p+\nu \Delta
{\bf v}+{\bf g}(z) \alpha T, \\
\textrm{div}{\bf v}=0, \\
\dfrac{\partial T}{\partial t}+({\bf v}\cdot \textrm{grad})T=k\Delta T,
\end{array}
\right.  t>0
\end{equation}
where $\nu$ is the coefficient of kinematic viscosity, $\rho$ the density, $\alpha$ the thermal expansion coefficient, $k$ the thermal
diffusitivity, $p$ is the pressure, $T$ the temperature, ${\bf v}$ the velocity and the gravity $g(z)$ is defined by ${\bf g}(z)= gH(z){\bf k}$,
where $ g$ is a constant.
\par The linear stability of the conduction stationary
solution characterized by ${\bf v}=0$ of equations (1), written in the nondimensional form, against normal mode perturbations, is governed by a
two-point problem for the ordinary differential equations
\begin{equation}
\left\{
\begin{array}{l}
(D^{2}-a^{2})^{2}W=RH(z)a^{2}\Theta,\\
(D^{2}-a^{2})\Theta=-RN(z)W,
\end{array}
\right.
\end{equation}
where $D=\dfrac{d}{dz}$, $R^{2}$ is the Rayleigh number, $a$ is the wavenumber and $W$ and $\Theta$ are the factors in $v$ and $\theta$
respectively, depending on $z$.
\par Consider $N(z)\equiv 1$ and $H(z)=1+\varepsilon h(z)$,
 $z\in (0,1)$. The parameter $\varepsilon$ represents a scale for $h(z)$. In this case,  the
 two-point problem for (2) consists of the ordinary differential equations \cite{Str}
\begin{equation}
\label{eq:CG_model} \left\{
\begin{array}{l}
(D^{2}-a^{2})^{2}W=R[1+\varepsilon h(z)]a^{2}\Theta,\\
(D^{2}-a^{2})\Theta=-RW
\end{array}
\right. \end{equation} and the usual boundary conditions for free boundaries read
\begin{equation}
\label{eq:CG_bc} W=D^{2}W=\Theta=0 \ \ \ \textrm{ at }z=0,1.
\end{equation}
 \par We look for the smallest eigenvalue $R$ (the Rayleigh number) in
(\ref{eq:CG_model})-(\ref{eq:CG_bc}) defining the neutral manifold.
\par In the case of rigid boundaries, i.e. the boundary conditions are
\begin{equation}
\label{eq:CG_bc_rigid}
W=DW=\Theta=0 \textrm { at } z=0,1
\end{equation}
 the eigenvalue problem has been studied by us in some previous papers, e.g.
\cite{Drag1}, \cite{Drag2}. The problem (\ref{eq:CG_model}) - (\ref{eq:CG_bc_rigid}) was also investigated in \cite{Str}. Straughan performed numerical evaluations of the Rayleigh number by using the energy method for some various gravity fields and our
numerical evaluations obtained in \cite{Drag1}, \cite{Drag2} were similar to those obtained in \cite{Str}. Herein, for the same varying gravity
fields we will obtain numerical results in the case of free boundaries.
\par In \cite{Her1}, \cite{Her2} it is proved that when $H(z)N(z)\geq 0$ across the layer the principle of exchange of stabilities
holds. In our case $N(z)\equiv 1$, $\varepsilon$ is a small parameter, for the chosen functions $h(z)$ the inequality $H(z)\geq 0$ is valid, therefore the principle of exchange of
stabilities holds no matter what the boundary conditions.  We can mention that there are cases in which this condition is not satisfied and the
principle of exchange of stabilities holds.
\par Although their study started in 1961,
the eigenvalue problems governed by systems of ordinary differential equations with variable coefficients and depending on many physical
parameters are difficult to solve. In most cases, an approximative solution for this type of problems can be obtained using spectral methods.
Since it has the advantage of optimal analysis, i.e. the analytical study is not a tedious one and the obtained numerical results are very good
compared to other methods,  the Galerkin method was chosen for the study of the eigenvalue problem (\ref{eq:CG_model})-(\ref{eq:CG_bc}).
\par The expansion sets of functions used for the various fields
encountered in the convection problems from hydrodynamic stability theory (e.g. the velocity field, the temperature field, the concentration
field) must have a  basic property: they must be easy to evaluate. That is why, most of the times, the trigonometric and the polynomial functions,
easy to evaluate, are used. A second property is the completeness of the expansion sets of functions. This assures that each function of the given
space can be written as a linear combination of functions from the considered set (or, more likely, as a limit of such a linear combination). The
Chebyshev polynomials, the Legendre polynomials, the Hermite functions, the sine and cosine functions, satisfy this
condition.

\par In \cite{Drag1} the two-point problem (\ref{eq:CG_model})-(\ref{eq:CG_bc}) was investigated using methods based on Fourier series of trigonometric functions.
 Herein, the trial and the test sets of functions will consists in Chebyshev and Legendre polynomials.
However, when the boundary conditions are very complicated the Galerkin approach is not easy to apply. That is way, in order to write the boundary conditions in a simpler form,
we introduced the function
 $\Psi=(D^{2}-a^{2})W$  and taking into account that $W=D^{2}W=0$ at $z=0,1$ we get  $\Psi=0$ at $z=0,1$.
\par By denoting ${\bf U}=(W, \Psi, \Theta)$ the eigenvector in (\ref{eq:CG_model})-(\ref{eq:CG_bc}), the two-point problem can be rewritten
 \begin{equation}
 \label{eq:CG_newmodel}
 \left\{
 \begin{array}{l}
 L_{1}{\bf U}=0\Leftrightarrow (D^{2}-a^{2})W-\Psi=0,\\
 L_{2}{\bf U}=0\Leftrightarrow (D^{2}-a^{2})\Psi-R(1+\epsilon h(z))a^{2}\Theta=0,\\
 L_{3}{\bf U}=0\Leftrightarrow (D^{2}-a^{2})\Theta+RW=0,
 \end{array}
 \right.
 \end{equation}
 with the boundary conditions
 \begin{equation}
 \label{eq:CG_bc_newmodel}
 W=\Psi=\Theta=0 \textrm{ at }z=0,1.
 \end{equation}
\section{Methods based on shifted polynomials}
In order to perform not only an analytical study, but also to obtain numerical evaluations for the Rayleigh number $R^{2}$, each of the functions
from the unknown eigenvector ${\bf U}$, is approximated by a truncated series of orthogonal polynomials, in our case Chebyshev and Legendre
polynomials. These polynomials are orthogonal on $[-1,1]$. Since in this problem the range is $[0,1]$, we will use shifted polynomials, orthogonal
on $[0,1]$, obtained from the original polynomials by a variable transformation.
 \par The Chebyshev polynomials were widely used in spectral methods for ordinary differential equations, e.g. \cite{Bo1},\cite{Gh}, \cite{Mas1}. Here we
 present only some basic properties of these polynomials necessary for our study.\par   The Chebyshev polynomials (of the first kind) of degree $n$, $T_{n}(z)$, are orthogonal on $[-1,1]$ with
 respect to the weight function $w(z)=\dfrac{1}{\sqrt{1-z^{2}}}$, i.e. $\ds\int_{-1}^{1}T_{n}(z)T_{m}(z)w(z)=\dfrac{\pi}{2}c_{n}\delta_{mn}$,
  $c_{n}=\left\{ \begin{array}{l}2, \textrm{ if } n=0,\\ 1, \textrm{ if } n\geq 1 \end{array}  \right.$.
  The shifted Chebyshev polynomials (of the first kind)(SCP) of degree $n$ on $(0,1)$, $T_{n}^{*}(z)$, are defined by the relation $T_{n}^{*}(z)=T_{n}(2z-1)$.
 The following orthogonality relation holds
 \begin{equation}
 \label{eq:CG_ort}
 \ds\int_{0}^{1}T_{n}^{*}(z)T_{m}^{*}(z)w^{*}(z)dz=\left\{
 \begin{array}{l}
 \dfrac{\pi}{2}c_{n}\delta_{nm}, \textrm{ if } i=j,\\
0, \textrm{ if } i\neq j, \end{array}
 \right.
 \end{equation}
 with respect to the weight function $w^{*}(z)=\dfrac{1}{z(1-z)}$.
 The recurrence relation between $T_{n}^{*}$ has the form
$$T_{n}^{*}(z)=2(2z-1)T_{n-1}^{*}(z)-T_{n-2}^{*}(z).$$
Similarly with Shen\cite{Sh2}, let us introduce $M_{1}=\{\Phi^{*}_{k}(z)\}_{k\in \mathbb{Z}}$, a the complete set of orthogonal functions in
$L^{2}(0,1)$, $\Phi^{*}_{k}(z)$ defined by
 \begin{equation}
 \label{eq:CG_funct}
 \Phi^{*}_{k}(z)=T_{k}^{*}(z)-T_{k+2}^{*}(z)
 \end{equation}
and satisfying boundary conditions of the type $\Phi^{*}_{k}(0)=\Phi^{*}_{k}(1)=0$. Then the unknown  functions  $\Psi$, $W$, $\Theta$ can be expanded
upon the complete set $M_{1}$ and they satisfy automatically all the boundary conditions. We have
 \begin{equation}
 \label{eq:CG_dezv}
 W=\sum\limits_{k=0}^{n}W_{k}\Phi_{k}^{*}(z),\ \ \Theta=\sum\limits_{k=0}^{n}\Theta_{k}\Phi_{k}^{*}(z), \ \
 \Psi=\sum\limits_{k=0}^{n}\Psi_{k}\Phi_{k}^{*}(z).
 \end{equation}
 \par The system (\ref{eq:CG_model}) can then be written in terms of the expansion functions only. Imposing the condition that left-hand side equations of
 the system to be orthogonal on $\Phi_{i}^{*}$, $i=0,1,...,n$, we get the algebraic system
 \begin{equation}
 \label{eq:CG_alg_sist}
\left\{
\begin{array}{l}
\sum\limits_{k=0}^{n}\Big\{\Big((D^{2}-a^{2})\Phi_{k}^{*}(z), \Phi_{i}^{*}(z)\Big)W_{k}-(\Phi_{k}^{*}(z), \Phi_{i}^{*}(z))\Psi_{k}\Big\}=0,\\
\sum\limits_{k=0}^{n}\Big\{\Big((D^{2}-a^{2})\Phi_{k}^{*}(z), \Phi_{i}^{*}(z)\Big)\Psi_{k}-Ra^{2}((1+\epsilon h(z))\Phi_{k}^{*}(z), \Phi_{i}^{*}(z))\Theta_{k}\Big\}=0,\\
\sum\limits_{k=0}^{n}\Big\{\Big((D^{2}-a^{2})\Phi_{k}^{*}(z), \Phi_{i}^{*}(z)\Big)\Theta_{k}+R(\Phi_{k}^{*}(z), \Phi_{i}^{*}(z))W_{k}\Big\}=0\\
\end{array}
\right.
 \end{equation}
in  the unknown coefficients  $W_{k}$, $\Psi_{k}$, $\Theta_{k}$. Since not all these coefficients are null,
the condition that the determinant of the system
vanish is imposed leading to the secular equation.
 \par Following \cite{Mas1} it is easy to deduce the following derivation formulae
 \begin{equation}
 \label{eq:CG_der1}
 (\Phi_{k}^{*}(z))'=2\Big\{2k\sum\limits_{{\tiny \begin{array}{l}
 r=0\\
 k-r \textrm{ odd }
 \end{array}}}^{k-1}T^{*}_{r}(z)-2(k+2)\sum\limits_{{\tiny \begin{array}{l}
 r=0\\
 k+2-r \textrm{ odd }
 \end{array}}}^{k+1}T^{*}_{r}(z)\Big\}
 \end{equation} for the first derivative of the function $\Phi_{k}^{*}$ and
\begin{equation}
 \label{eq:CG_der2}
 (\Phi_{k}^{*}(z))''=4\Big\{\sum\limits_{{\tiny \begin{array}{l}
 r=0\\
 k-r \textrm{ even }
 \end{array}}}^{k-2}(k-r)k(k+r)T^{*}_{r}(z)-\sum\limits_{{\tiny \begin{array}{l}
 r=0\\
 k+2-r \textrm{ even }
 \end{array}}}^{k}(k+2-r)(k+2)(k+2+r)T^{*}_{r}(z)\Big\}
 \end{equation}
 for the second one. In the numerical evaluations we will take into account that the first term in each of the involved sums is halved.
\par The presence of the varying gravity field led to nonconstant coefficients, such that the analytical expression
of the scalar product $(h(z)\Phi_{k}^{*}(z), \Phi_{i}^{*}(z))$ is based on the relation \cite{Gh}
 \begin{equation}
 \label{eq:CG_putere}
 z^{r}T_{s}(z)=\dfrac{1}{2^{r}}\sum\limits_{i=0}^{r}C_{r}^{i}T_{s-r+2i}(z).
 \end{equation}
The analytical expressions of all the other scalar products from (\ref{eq:CG_alg_sist}) are taking with respect to the weight function $w^{*}(z)$ and
deduced by taking into account the orthogonality
relation (\ref{eq:CG_ort}).
\par The secular equation leading to the neutral values of the Rayleigh number can be deduced in a similar way by using shifted Legendre
polynomials.
 \par   Let
  $$H_{0}^{1}(0,1)=\{f| f, f'\in L^{2}(0,1), f(0)=f(1)=0 \}, $$
be a Hilbert space and denote by $L_{k}$ the Legendre polynomials defined on $(-1,1)$. Then the shifted Legendre polynomials $Q$ (SLP) on $(0,1)$ are defined by
the relation $Q_{k}(x)=L_{k}(2x-1)$ and they are orthogonal on the interval $(0,1)$, i.e. $
\ds\int_{0}^{1}Q_{i}Q_{j}dz=\dfrac{1}{2i+1}\delta_{ij}$. Using the identity \cite{hStr1}
\begin{equation}
\label{eq:rec_der} 2(2i+1)Q_{i}(z)=Q'_{i+1}(z)-Q'_{i-1}(z),
\end{equation}
we define the set $M_{2}$ of orthogonal functions $\phi_{i}$,
$$\phi_{i}(z)=\ds\int_{0}^{z}Q_{i}(t)dt=\dfrac{Q_{i+1}-Q_{i-1}}{2(2i+1)}, i=1,2,...$$
 that satisfy boundary conditions of the type $\phi_{i}(0)=\phi_{i}(1)=0$ at $z=0$ and
$1$ such that the set $M_{2}$ is complete in $H_{0}^{1}(0,1)$.
\par Therefore we can write the unknown functions  $W$, $\Psi$, $\Theta$ as series
in the form
\begin{equation}
\label{eq:dezv}  W=\sum\limits_{i=1}^{n} W_{i}\phi_{i}(z), \  \  \Psi=\sum\limits_{i=1}^{n} \Psi_{i}\phi_{i}(z),\ \
\Theta=\sum\limits_{i=1}^{n}\Theta_{i}\phi_{i}(z).
\end{equation}
The secular equation is obtained following the same steps in the analytical study as before, i.e.
\begin{equation}
\label{eq: CG_ec_sec}
\begin{tabular}{|ccc|}
$((D^{2}-a^{2})\phi_{i},\phi_{k})$& $-1$&$0$\\
& &\\
$0$ &$((D^{2}-a^{2})\phi_{i},\phi_{k})$ & $-Ra^{2} ((1+\epsilon h(z))\phi_{i},\phi_{k})$\\
& & \\
$((D^{2}-a^{2})\phi_{i}, \phi_{k})$&$0$&$((D^{2}-a^{2})\phi_{i},\phi_{k})$
\end{tabular}=0.
\end{equation}
 \section{Numerical evaluations}
 The numerical evaluations were obtained for different significant values of the scale parameter $\epsilon$ and the wavenumber $a$. The number of functions in
 the expansion sets was small ($n=4$), but the obtained approximative values of the Rayleigh number were similar to ones obtained with other methods and we considered that
 a small improvement obtained for $n>4$ would not justify
more time. In order to compare our results and implicitly to test the method, we  took into consideration three variable gravity fields from
\cite{Str}, i.e. $h(z)=-z$, $h(z)=-z^{2}$ and $h(z)=z^{2}-2z$. The numerical results presented in Tables 1,2,3 show that a decreasing gravity
field enlarge the domain of stability. For $\epsilon=0$, we obtained similar evaluations with the classical ones from \cite{Ch}.
 \begin{center}
 \begin{tabular}{|c|c|c|c|}
 \hline
 $\epsilon$&$a^2$&$R^{2}-SCP$&$R^{2}-SLP$\\
 \hline
 $0.0$&$4.92$&$657.512$&$675.05$\\
 \hline
 $0.01$&$4.92$&$660.747$&$678.45$\\
 \hline
 $0.03$&$4.92$&$667.653$&$685.33$\\
 \hline
 $0.33$&$4.92$&$787.363$&$808.303$\\
 \hline
 $0.2$&$5.00$&$730.459$&$749.95$\\
 \hline
 $0.2$&$9.00$&$829.44$&$846.70$\\
 \hline
 $0.5$&$7.5$&$930.982$&$952.07$\\
 \hline
 $0.5$&$9.00$&$994.393$&$1015.27$\\
 \hline
 $0.75$&$10.0$&$1251.178$&$1276.05$\\
 \hline
 \end{tabular}
 \end{center}
 \begin{center}
 {\small {\bf Table 1.} Numerical values of the Rayleigh number for various values of the parameters for $h(z)=-z$.}
 \end{center}
 \begin{center}
 \begin{tabular}{|c|c|c|c|}
 \hline
 $\epsilon$&$a^2$&$R^{2}-SCP$&$R^{2}-SLP$\\
 \hline
 $0.0$&$4.92$&$657.512$&$675.05$\\
 \hline
 $0.01$&$4.92$&$659.41$&$676.99$\\
 \hline
 $0.03$&$4.92$&$663.17$&$680.90$\\
 \hline
 $0.33$&$4.92$&$725.06$&$745.21$\\
 \hline
 $0.2$&$5.00$&$696.80$&$715.87$\\
 \hline
 $0.2$&$9.00$&$791.24$&$808.22$\\
 \hline
 $0.5$&$7.5$&$813.28$&$833.21$\\
 \hline
 $0.5$&$9.00$&$868.72$&$888.54$\\
 \hline
 $0.75$&$10.0$&$993.51$&$1016.20$\\
 \hline
 \end{tabular}
 \end{center}
 \begin{center}
 {\small {\bf Table 2.} Numerical values of the Rayleigh number for various values of the parameters for $h(z)=-z^{2}$.}
 \end{center}
 \begin{center}
 \begin{tabular}{|c|c|c|c|}
 \hline
 $\epsilon$&$a^2$&$R^{2}-SCP$&$R^{2}-SLP$\\
 \hline
 $0.0$&$4.92$&$657.512$&$675.05$\\
 \hline
 $0.01$&$4.92$&$662.29$&$679.91$\\
 \hline
 $0.03$&$4.92$&$671.95$&$689.83$\\
 \hline
 $0.33$&$4.92$&$861.25$&$882.05$\\
 \hline
 $0.2$&$5.00$&$767.40$&$787.44$\\
 \hline
 $0.2$&$9.00$&$871.37$&$889.03$\\
 \hline
 $0.5$&$7.5$&$1088.2$&$1110.46$\\
 \hline
 $0.5$&$9.00$&$1162.4$&$1184.11$\\
 \hline
 $0.75$&$10.0$&$1687.8$&$1713.45$\\
 \hline
 \end{tabular}
 \end{center}
 \begin{center}
 {\small {\bf Table 3.} Numerical values of the Rayleigh number for various values of the parameters for $h(z)=z^{2}-2z$.}
 \end{center}
 \section{Conclusions}
 In this paper we presented two shifted polynomials-based methods for the study of the linear stability of
the mechanical equilibrium of a horizontal layer of a viscous incompressible fluid heated from below in the case of a variable gravity field.
\par We provided numerical results for various gravity fields, decreasing but not all linear. These results proved to agree quite well
with the results obtained by us with the Galerkin method based on trigonometric functions \cite{Drag1} and also with the classical one existing in
the literature. The numerical evaluations of the Rayleigh number were obtained for different values of the physical parameters, allowing a
conclusion on the effects of these parameters on the stability domain. It was seen that the domain of stability decreases as the gravity field is
increasing.
\par Although both expansion sets of functions led to good numerical evaluations of the Rayleigh number, it seems like the method based on the expansion upon shifted Chebyshev
polynomials was more effective.  However, in most cases,  the Legendre polynomials are preferred in the Galerkin approach and the Chebyshev polynomials are
considered suitable for the collocation methods.
 
\end{document}